\documentclass[twocolumn,prl,floatfix,showpacs,superscriptaddress]{revtex4-1} %

\usepackage{color,amsmath, amssymb, graphicx, amsfonts ,  setspace,epstopdf}

\newcommand{\dif}{\ensuremath{\,\mathrm{d}}} 
\newcommand{\dro}[2]{\ensuremath{\frac{\dif{#1}}{\dif {#2}}}}

\begin{document}

\title{ Quantum critical origin of the superconducting dome in SrTiO$_3$}

\author{Jonathan M. Edge}
\affiliation{Nordita, KTH Royal Institute of Technology and Stockholm
  University, Roslagstullsbacken 23 106 91 Stockholm, Sweden}
\author{Yaron Kedem}
\affiliation{Nordita, KTH Royal Institute of Technology and Stockholm
  University, Roslagstullsbacken 23 106 91 Stockholm, Sweden}
\author{Ulrich Aschauer}
\affiliation{Materials Theory, ETH Zurich, Wolfgang-Pauli-Strasse 27, CH-8093 Z\"urich, Switzerland}
\author{Nicola A. Spaldin}
\affiliation{Materials Theory, ETH Zurich, Wolfgang-Pauli-Strasse 27, CH-8093 Z\"urich, Switzerland}
\author{Alexander V. Balatsky}
\affiliation{Institute for Materials Science, Los Alamos National Laboratory, Los Alamos,
NM, 87545, USA}
\affiliation{Nordita, KTH Royal Institute of Technology and Stockholm University, Roslagstullsbacken 23
106 91 Stockholm, Sweden}

\date{\today}
\begin{abstract}
We investigate the origin of superconductivity in doped SrTiO$_3$ (STO) using a combination of density functional and strong coupling theories within the framework of quantum criticality. Our density functional calculations of the ferroelectric soft mode frequency as a function of doping reveal a crossover from quantum paraelectric to ferroelectric behavior at a doping level coincident with the experimentally observed top of the superconducting dome. Based on this finding, we explore a model in which the superconductivity in STO is enabled by its proximity to the ferroelectric quantum critical point and the soft mode fluctuations provide the pairing interaction on introduction of carriers. Within our model, the low doping limit of the superconducting dome is explained by the emergence of the Fermi surface, and the high doping limit by departure from the quantum critical regime. We predict that the highest critical temperature will increase and shift to lower carrier doping with increasing $^{18}$O isotope substitution, a scenario that is experimentally verifiable. 
\end{abstract}

\maketitle

Strontium titanate (STO) is a cubic perovskite with the ideal prototype structure at room temperature and a tetragonal
structure below $\sim$100K due to symmetry-lowering antiferrodistortive (AFD) rotations of the TiO$_6$ octahedra.
It is characterized by a number of remarkable properties. It was the first superconducting oxide to be discovered and shows a dome as a function of doping, similar to that of the high-$T_c$ cuprates\cite{Koonce1967}, but  
with its maximum transition temperature at $~T_c \simeq 0.4$K. Early tunneling measurements \cite{Binnig1980}
and subsequent experiments \cite{Lin2013} 
suggested an unusual two-band superconductivity, consistent with the closely spaced lowest conduction bands 
at the center of the Brillouin zone. In addition, the onset of superconductivity has been shown to occur
at remarkably low carrier concentrations of $~10^{18}$e/cm$^3$\, \cite{Lin2013}. 
Despite a long-running interest in its origin \cite{Koonce1967}, a complete theoretical account of the superconducting dome 
remains elusive, and many aspects of superconductivity in STO remain a puzzle.

The dielectric behavior of STO is also unusual. The dielectric constant is strongly temperature dependent, and diverges at low
temperature in a manner characteristic of a ferroelectric phase transition \cite{Muller1979}. 
Rather than manifesting ferroelectric
behavior, however, STO is a so-called quantum paraelectric, in which quantum fluctuations at zero temperature suppress the 
transition to the ferroelectric state \cite{Muller1979}. The quantum paraelectric state is characterized by low energy excitations 
and large ferroelectric fluctuations  \cite{suzuki2013}, and it has been speculated that these might
be relevant for the superconductivity \cite{BussmannHolder/Simon/Buttner:1989,deLima_et_al:2015}. 
Indeed, early descriptions \cite{Koonce1967,Appel1969} of the superconducting dome in STO were based on the effects of screening of 
the interaction between electrons and the optical phonons responsible for the large dielectric response. 
Because heavier $^{18}$O atoms suppress the quantum fluctuations, STO develops ferroelectric order on isotope 
substitution of $^{16}$O with $^{18}$O, and the composition with 35\% $^{18}$O substitution was recently reported to be 
a ferroelectric quantum critical point (QCP) \cite{Rowley2014}.

We present a model in which these two features -- proximity to the ferroelectric QCP and the unusual superconducting properties 
-- are intimately related, and the superconducting dome emerges as a result of the quantum critical ferroelectric fluctuations.
A connection between the formation of a superconducting dome and quantum criticality has been extensively discussed in the context of 
unconventional superconductivity, both in heavy fermion materials and in the cuprates \cite{Gegenwart2008,Sachdev2000,Sebastian2010}. 
It is proposed that competing phases close to the quantum critical point lead to low energy excitations such that
any residual interactions drive the system to a new, possibly superconducting phase. 
In heavy fermion materials and the cuprates a magnetic quantum critical point with associated 
spin excitations has been invoked to explain superconductivity.
In STO the elementary
excitations associated with the ferroelectric quantum critical point are optical phonon modes. 
As a result we expect differences in the nature of the superconducting order: Magnetic fluctuations typically produce unconventional 
superconducting order such as $d$-wave singlets for antiferromagnetic fluctuations \cite{Monthoux1991,ScalapinoRMP} or $p$-wave 
triplet states for ferromagnetic fluctuations \cite{Roussev2001}. The ferroelectric fluctuations in STO, in contrast, involve 
$q = 0$ phonon modes and as such are candidates for pairing interactions that introduce conventional $s$-wave superconducting order, as observed in experiments \cite{Lin2015}. 

\begin{figure}[tb]
\centering
\includegraphics[width=\columnwidth]{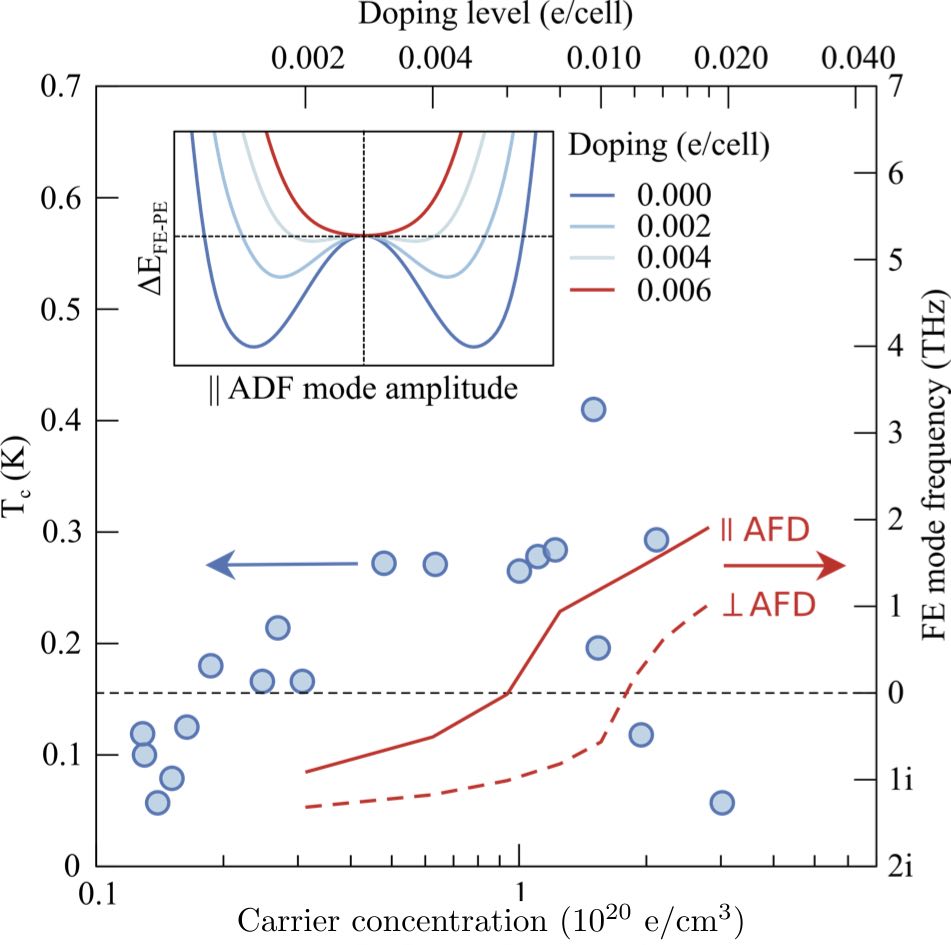}
\caption{Literature values of the superconducting critical temperature \cite{Koonce1967}  (circles) and calculated frequencies (this work) of the ferroelectric modes parallel- ($\parallel$) and 
perpendicular ($\perp$) to the axis of the AFD rotations 
(red solid and dashed lines) as a function of the carrier concentration. The imaginary frequencies obtained
at low doping indicate negative restoring forces corresponding to ferroelectric instabilities; as the carrier concentration is increased 
the ferroelectric mode hardens and its phonon frequency becomes real. The inset shows the calculated energy as a function of 
ferroelectric mode amplitude for various doping levels, illustrating the crossover from the classic ferroelectric double well
potential energy to a single well, indicating a paraelectric ground state on increasing doping. 
As the charge carrier concentration is increased, $T_c$ first increases and then decreases, forming the characteristic superconducting
dome. We see that the doping concentration at which $T_c$ drops to zero, $\sim10^{20}$e/cm$^3$, closely matches that at which
the ferroelectric mode hardens.
}
\label{fig:freq_of_FE-mode_and_Tc_vs_doping}
\end{figure}

Our model is motivated by our density functional theory (DFT) calculations of the zone-center ($q=0$) soft-mode optical phonon 
frequency as a function of electron doping, which shows an intriguing correlation with experimental measurements of the 
superconducting dome. This ferroelectric soft mode, which consists predominantly of opposite Ti cation and O anion displacements (for details, see Ref. \onlinecite{Aschauer2014}), 
has a calculated imaginary frequency at zero doping, indicating the presence of a ferroelectric instability.
The calculated potential energy as a function of the relative position of anions and cations 
(see supplementary material Sec. I for details)
shows the characteristic double well form, with the two minima corresponding to ferroelectric structural ground states with
opposite polarizations. 
In practice, quantum fluctuations between the two wells suppress the ferroelectricity in STO, and give it its quantum 
paraelectric behavior. 
In Fig.~\ref{fig:freq_of_FE-mode_and_Tc_vs_doping} we show how, on electron doping, the modulus of the mode frequency decreases, corresponding to a weakening of the ferroelectric instability, and the frequency 
eventually becomes real -- signaling a single high-symmetry energy minimum (inset to Fig.~\ref{fig:freq_of_FE-mode_and_Tc_vs_doping}) 
-- at a doping concentration of $\sim10^{20}$ cm$^{-3}$. Since there is now only one minimum of the potential well, there are clearly no
quantum fluctuations between equivalent states.
At the same doping level, the experimentally measured superconducting transition temperature starts to reduce.
Since soft modes are characteristic of quantum criticality \cite{Sachdev2011}, we propose therefore
the following model for the superconducting dome in STO:  
First, superconductivity is favored when the quantum fluctuations favored by the soft lattice modes increase the
superconducting coupling constant $\lambda$. However, these are strongest at low doping, where there are insufficient 
carriers to provide robust superconductivity. Increasing the doping level has the side-effect of reducing $\lambda$, which in turn determines the upper bound of the
superconducting dome. 

\begin{figure}[tb]
\centering
\includegraphics[width=\columnwidth]{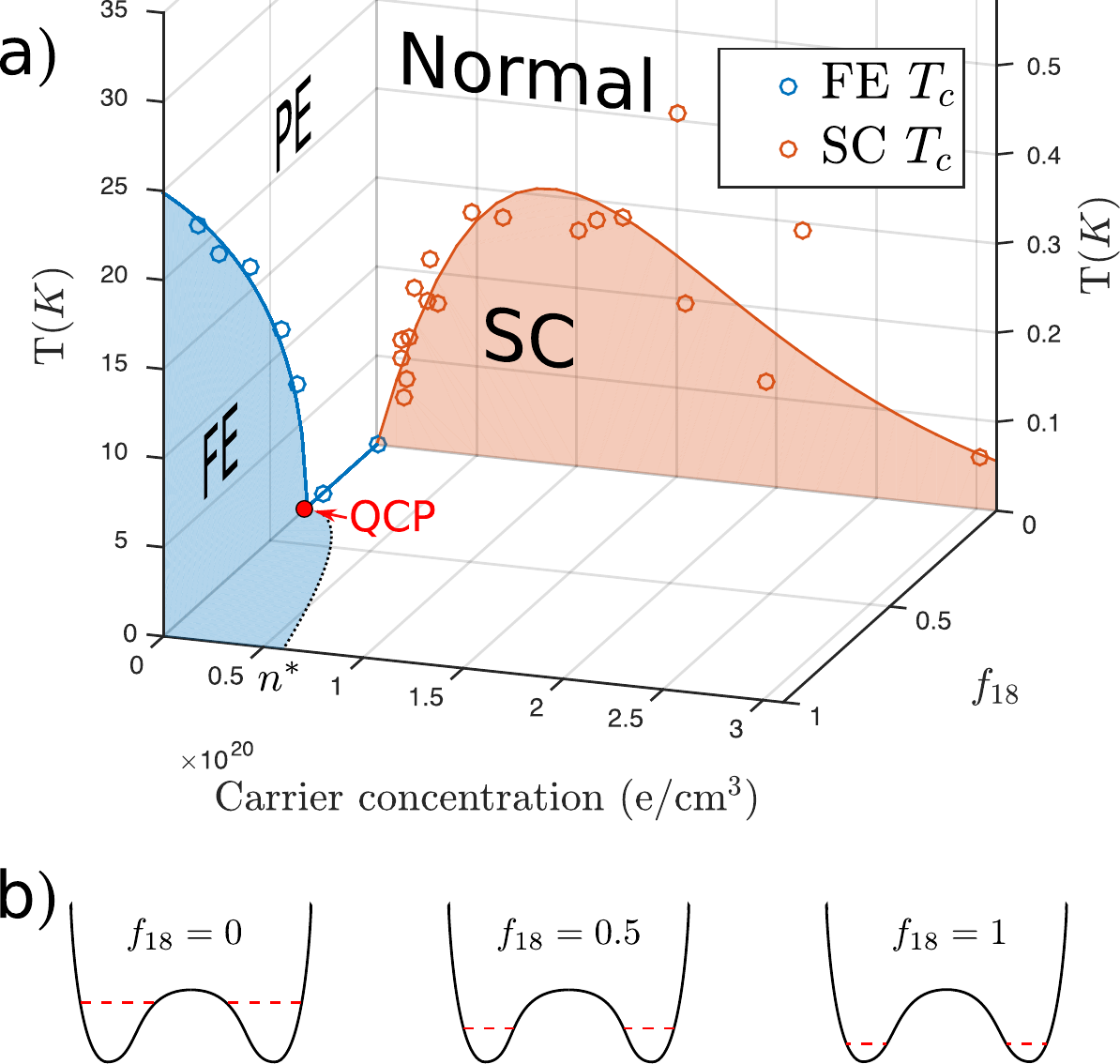}
\caption{a) Schematic phase diagram of STO as a function of carrier doping and isotope replacement. The orange circles mark the 
experimentally measured transition to superconductivity, as observed in Ref.~\onlinecite{Koonce1967}. The blue circles are the 
measured transition temperatures \cite{Wang2001}  from the paraelectric (PE)  to the ferroelectric (FE) phase as a function of
$^{18}$O isotope substitution. 
Our DFT calculations suggest that the ferroelectric phase penetrates slightly into the non-zero doping regime, but then quickly 
disappears as doping suppresses ferroelectricity, although no experimental data for this transition line is available. The maximal value of doping at which the ferroelectric phase persists is labeled as $n^*$. Although we have no precise calculation for $n^*$, its value should lie in the range $10^{19}<n^*<10^{20}$.
b) Schematic illustration for the lowering of the lowest energy levels (dashed red lines) in the double well potential (black solid line) as $f_{18}$ is increased.
}
\label{fig:schematic_plot_coupling}
\end{figure}

To test this hypothesis we propose 
isotopic substitution of $^{16}$O with $^{18}$O, which lowers the energies of the 
zero-point energy levels in the two minima and reduces the probability of tunneling between them
eventually favoring a ferroelectric ground state, see Fig.~\ref{fig:schematic_plot_coupling}b.
We know that at zero carrier doping, the paraelectric to ferroelectric transition occurs at about 
35\% $^{18}$O substitution and is a quantum critical point\cite{Rowley2014}.
In addition, our DFT calculations tell us that doping suppresses FE, and so the QCP 
should move to higher $^{18}$O fractions as doping is increased, implying the existence of a quantum critical line (QCL). This
allows us to construct the schematic phase diagram in Fig.~\ref{fig:schematic_plot_coupling}. 
Our DFT calculations give an upper bound for this QCL, which is the doping level at which the frequency of 
the FE mode becomes real and the quantum fluctuations are completely suppressed; in practice this represents the 
limit of infinitely heavy oxygen atoms and the actual critical transition will occur at much lower doping. 
Note that, at least at low $^{18}$O concentration, charge carriers only appear as one moves away from the quantum critical point, 
so the QCP is in fact located outside 
the superconducting dome. This is in contrast to the emergence of superconductivity in other systems close to a QCP, such as
the cuprate superconductors, in which the dome is approximately centered on the QCP. In those cases, the QCP occurs at 
substantial doping, where charge carriers are already available.

We quantify our proposed model by calculating $T_c$, assuming the scenario of soft critical modes in the limit of low 
doping. We first write a quantum model for the ferroelectric phase transition which yields a spectrum for the FE phonons. Then, 
we calculate the superconducting coupling constant, using the Macmillan formula \cite{McMillan1968}.

We use the order-disorder approach \cite{Yamada2004} to model the ferroelectric fluctuations of the modes shown in Fig.~\ref{fig:freq_of_FE-mode_and_Tc_vs_doping}.
We assume that these modes have Ising character.
By analogy with magnetic phase transitions, the transverse Ising model 
\begin{equation} \label{h}
H = \Gamma\sum_i \sigma_x(i) - \sum_{i,j} J_{i,j}\sigma_z (i) \sigma_z (j)
\end{equation}
can be used to describe the FE transition \cite{suzuki2013}.
Here $\sigma_{x,z}(i) $ are the Pauli matrices for site $i$, $\Gamma / \hbar $ is the onsite tunneling rate, $ J_{i,j}$ is the inter-site 
coupling, given by the energy difference between two cells with their dipoles aligned parallel or anti-parallel to each
other, and the eigenstates of $\sigma_{z} $ represent the state of the system in one of the two wells. 
The quantum phase transition occurs when $\Gamma \sim \sum_{j} J_{0,j} $  \cite{suzuki2013}.
Our DFT study shows that
doping the system will reduce the barrier and thus increase $\Gamma$. 
The excitations of (\ref{h}), in the paraelectric phase $ \Gamma > \sum_{j} J_{0,j}$, are given by  \cite{suzuki2013}
\begin{equation} \label{w}
\omega^2_\mathbf{q} = 4 \Gamma  \left( \Gamma - \langle \sigma_x \rangle J_\mathbf{q} \right)
\end{equation}
where $J_\mathbf{q} = \sum_j J_{0,j} e^{i \mathbf{R_j} \mathbf{q} } $ is the Fourier transform of the coupling and 
$\langle \sigma_x \rangle \sim 1 $ is the average of $ \sigma_x(i)$. 
In our analysis we consider only nearest-neighbor coupling for simplicity. Long range interactions make the calculation more intricate but do not yield any qualitative changes. Furthermore,
since the antiferrodistortive rotations of the TiO$_6$ octahedra render the lattice highly anisotropic, we treat 
the system as one dimensional. Thus, we write the coupling as $J_\mathbf{q} = 2 J \cos(q) $, where $J$ is a constant 
and $q$ is the wave number in the direction of the largest coupling.

When the system is close to the phase transition it becomes gapless as the lowest excitation 
softens, $\omega_{\mathbf{q}=0} \rightarrow 0$ (see supplementary material, Sec. II)). This is accompanied a large 
susceptibility and an enhanced electron-phonon coupling.
To
quantify this idea we calculate the dependence of $T_c$ on the phononic spectrum using the formalism of Eliashberg 
strong-coupling theory.
The coupling constant for superconductivity is given by \cite{McMillan1968}
\begin{equation} \label{lambda}
\lambda=\int_ 0^{\infty} \alpha^2(\omega) F(\omega) {d\omega \over \omega}
\end{equation}
where $\alpha(\omega)$ is the electron-phonon coupling, which we assume to be the constant $\alpha$, and $F(\omega)$ is the spectral 
density of the phonons. In the limit of a van Hove singularity at $q=0$, so that $F(\omega) \sim \delta(\omega - \omega_0) $, this
yields 
\begin{equation} \label{eq:lambda_from_delta_fn_omega_expr}
\lambda= \alpha^2 {1 \over \omega_{\mathbf{q}=0} (f_{18}, E_F) } \quad , 
\end{equation}
which already captures the main physical picture of soft-mode enhanced superconductivity. The full solution is obtained by 
inserting $F(\omega) = \int dq \delta(\omega - \omega_q) $ into (\ref{lambda}) and transforming it to an integral over $q$: 
$\lambda= \int \alpha^2 {dq \over \omega_q}$, where $\omega_q$ is given by (\ref{w}). One then obtains
\begin{equation}
\lambda \sim  \int_{-\pi}^\pi {dq \over 2 \Gamma \sqrt{1 - 2 J \cos(q)  / \Gamma} } \quad.
\end{equation}

The critical temperature can then be obtained by combining this coupling constant with the standard expression
(see for example Ref.~\onlinecite{AGD-QFT_methods_1965})
\begin{equation} \label{Tc1}
1 ={\lambda \over 2 \pi^2} \int_{-E_F}^0  d \epsilon N(\epsilon) {\tanh \left( \epsilon / 2 T_c \right) \over \epsilon }
\end{equation}
where $\epsilon$ is the energy relative to the Fermi energy, $E_F$, and $N(\epsilon)$ is the density of states. 
The lower limit of the integral is set by $N(\epsilon)=0$ 
at and below the bottom of the band where $\epsilon < - E_F$. The upper limit is set by the Fermi level, where we
define $\epsilon = 0$. 
Since in the low doping scenario that we consider here the relevant energy range is close to the bottom of the band, 
we can assume that  $N(\epsilon ) \sim \sqrt{\epsilon + E_F} $
close to $\epsilon=-E_F$. Using $x=\epsilon/T_c$ equation (\ref{Tc1}) then becomes
\begin{equation} \label{numInt}
{D \over \lambda} =  \sqrt{T_c} \int_{- E_F /T_c}^{0} dx \sqrt{x+ E_F /T_c } {\tanh (x/2) \over x},
\end{equation}
where $D$ is a constant of proportionality. Note that $T_c$ has a double dependence on $E_F$: one directly from the limit of the
integral in (\ref{numInt}) 
and the other from the dependence of $\lambda$ on the tunneling rate $\Gamma$ on $E_F$ through its dependence on the carrier concentration.

\begin{figure}
\centering
\includegraphics[width=\columnwidth]{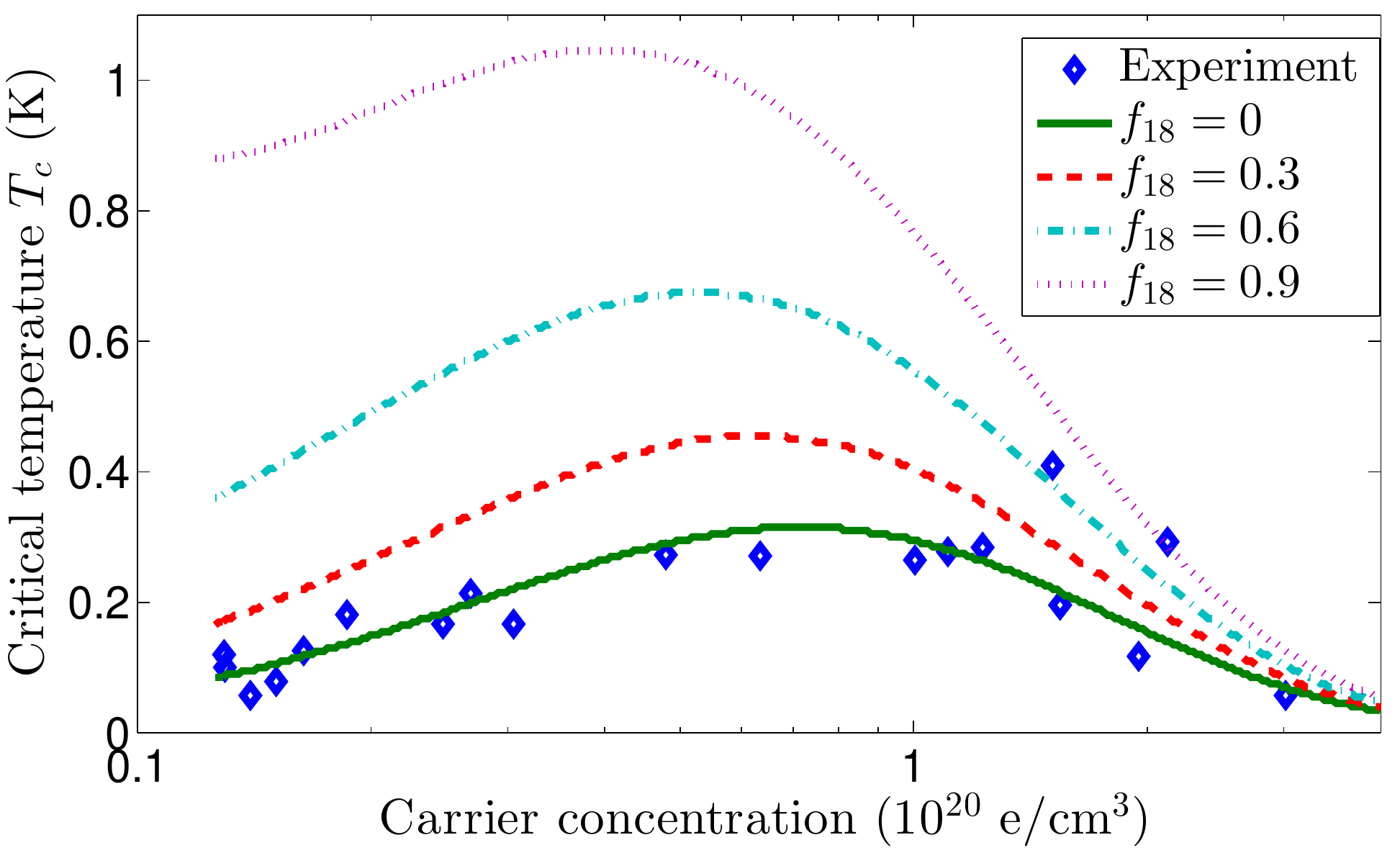}
\caption{Calculated $T_c$ as a function of doping level for several fractions of isotope replacement, $f_{18}$.
  The blue diamonds are experimental results taken from Ref. \cite{Koonce1967}.
  Replacing $^{16}$O with $^{18}$O moves the QCP closer to the doping range relevant for superconductivity and causes a significant enhancement in $T_c$. 
we use the parameters $A=0.4$, $B=10^{-6}$ K$^{-2}$, $C=2.5 \times 10^{-3}$ K$^{-1}$, $D=95$ K$^{1/2}$, as defined in the main text.
}
\label{fig:TcDomel}
\end{figure}

Before we can solve Eq.~\ref{numInt} numerically to obtain $T_c$ as a function of $E_F$, we need the explicit dependence of the parameters 
of our model (\ref{h}) on doping and isotope replacement. The quantity with the largest quantitative influence is the ratio 
$\Gamma / 2J$, which is equal to one on the QCL. For simplicity we set $2J = 1$ and consider only the dependence of $\Gamma$. 
As discussed above, carrier doping decreases the barrier between the two wells and thus increases the tunneling energy $\Gamma$. 
This effect starts at low doping level and becomes very strong around carrier concentrations of $10^{20}$e/cm$^3$, so we consider 
both a linear and quadratic dependence on $E_F$. $^{18}$O replacement on the other hand should decrease $\Gamma$ approximately
linearly as the zero-point energy levels shift deeper into the wells. 
Furthermore, we require that at zero doping and 35\% $^{18}$O substitution, which is the known QCP, $\Gamma$ 
should equal unity. Thus we use the form
\begin{equation}
  \Gamma = 1 - A (f_{18} - 0.35) + B E_F^2 + C E_F,
  \label{expansion_gamma}
\end{equation}
where $f_{18}$ is the $^{18}$O fraction and the constants $A$, $B$, and $C$ are chosen so that the calculated $T_c$ for $f_{18}=0$ matches
the experimental value.  We then use the expression $\Gamma$ from Eq.~\eqref{expansion_gamma} and insert this into Eq.~\eqref{numInt} to calculate $T_c$.

In Fig. \ref{fig:TcDomel} we plot our calculated $T_c$ as a function of the Fermi energy (converted to carrier concentration)
for various values of $f_{18}$. 
Two features are clear from the plot: i), we find a significant enhancement of $T_c$ with increased $^{18}$O content, reflecting
the fact that the isotope substituted system is closer to the QCP. ii), we find that the peak of the superconducting dome
shifts to lower carrier concentrations, since the enhancement of $\lambda$ and thus $T_c$ is strongest close to the QCP, as can be seen from Eq.~\eqref{lambda}. We note that, even when $f_{18}$
exceeds 0.35, doping quickly reduces the depth of the double wells, allowing quantum fluctuations to return STO to the 
quantum paraelectric state. Thus, apart from the limit of very low doping, all systems we consider have paraelectric, 
not ferroelectric ground states.
In our mechanism for superconductivity in STO, increasing the atomic mass leads to an increase of the critical temperature. That is $\dro{T_c}{f_{18}}/T_c>0$ (for details, see supplementary material, Sec. III). This differs profoundly from the well-known isotope effect in BCS superconductors, in which $\frac{\Delta T_c}{T_c}=-\frac12 \frac{\Delta M}M$ \cite{isotope}, where $M$ is the mass of the atoms. This arises from the dependence of $T_c$ on the Debye frequency.

We have provided a description of the superconducting dome in STO in which the QCP at zero doping provides low energy soft 
phonon excitations, which lead to a large coupling constant. Increasing the doping provides carriers for superconductivity but reduces 
the ferroelectric quantum fluctuations and decreases the coupling constant, eventually suppressing the superconductivity and limiting 
the top of the superconducting dome. Since isotope substitution allows tuning of the QCP, our model predicts a large and unusual isotope 
effect on $T_c$, see Fig.~\ref{fig:TcDomel}, which should be experimentally observable.
The understanding of the competition between carrier concentration and proximity to a QCP developed here provides a new design guideline 
in the search for novel superconducting compounds and suggests a route to engineering materials with higher $T_c$s through tuning the
location of their QCP. 

{\em Acknowledgments} We are grateful to D. Abergel, K. Behnia, J. Haraldsen, R. Fernandes, S. Raghu and P. W\"olfle for useful discussions. This work was supported by US DOE BES E304,
by the ETH Z\"urich (NAS and UA) and by the ERC Advanced Grant Program, No.\ 291151 (NAS and UA), No321031, KAW and LDRD (AVB and YK).

\appendix

\begin{widetext}

  \section{Supplemental material}

  \section{ Ferroelectric modes}
The ferroelectric mode frequencies plotted in Fig.~1 in the main text refer to the mode parallel to the axis of the antiferrodistortive rotations of the TiO$_6$ octahedra (solid line) and to the doubly degenerate perpendicular modes perpendicular to the octahedral rotations (dashed line). The displacements of the atoms within the unit cell for each of these modes are shown in Fig.~\ref{fig:FE-modes}.

The mode frequencies were computed using the frozen phonon approach as implemented in Phonopy \cite{Togo:2008jt} with forces obtained from density functional theory with the PBEsol functional \cite{Perdew:2008fa} as implemented in the Vienna \textit{ab initio} simulation package (VASP) \cite{Kresse:1993ty, Kresse:1994us, Kresse:1996vk, Kresse:1996vf}. In these force calculations, wavefunctions were expanded in plane waves up to a kinetic energy cutoff of 550 eV for PAW potentials \cite{Blochl:1994uk, Kresse:1999wc} with Sr(4s, 4p, 5s), Ti(3p, 3d, 4s) and O(2s, 2p) states in the valence. Reciprocal space was sampled using a 4x4x4 Monkhorst-Pack mesh \cite{Monkhorst:1976ta} for the 2x2x2 supercell used to describe the structure containing the fully relaxed AFD. Electron doping was performed by adding extra charge carriers to the cell while applying a neutralizing background charge.

\begin{figure}[tb]
\centering
\includegraphics[width=\columnwidth]{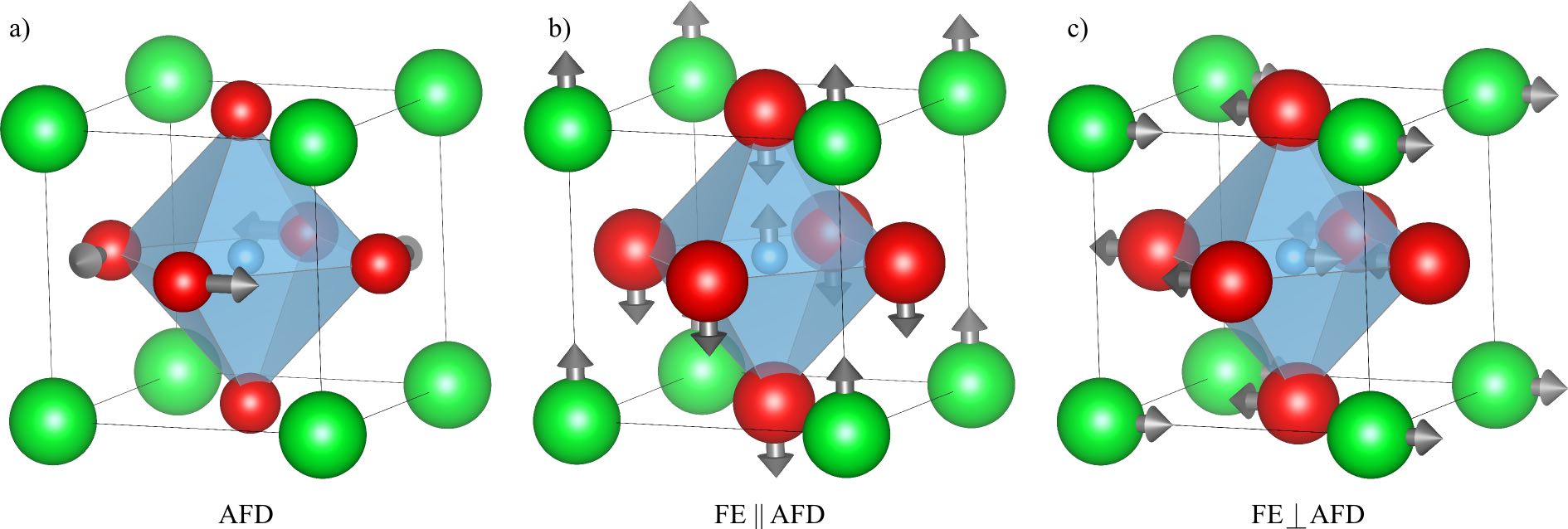}
\caption{The displacements of the atoms within the unit cell for a) the AFD mode as well as b) the FE mode parallel to the AFD axis and c) the FE mode perpendicular to the AFD axis.
}
\label{fig:FE-modes}
\end{figure}

\section{The frequency of the soft modes}

Our model predicts some unusual effect with regards to isotope replacement. One of them is the softening of the lowest frequency mode.
This frequency, $\omega_\mathbf{q=0}$, is given by Eq.~(2) in the main text. By using our model for the dependence of the tunneling energy on the $^{18}$O fraction (Eq.~(8) in the main text) we can explicitly calculate the change of this frequency with respect to the $^{18}$O fraction ($f_{18}$):
\begin{equation} \label{w}
{d \omega_\mathbf{q=0} \over d f_{18} } = - A { 2 \Gamma - 1 \over \sqrt{ \Gamma ( \Gamma -1) }},
\end{equation}
For $\Gamma>1$ this thus always lowers the mode frequency as the $^{18}$O fraction is increased, taking the system closer to the QCP.

\section{Equation for the change of $T_c$}

\begin{figure}[t]
\centering
\includegraphics[width=0.75\columnwidth]{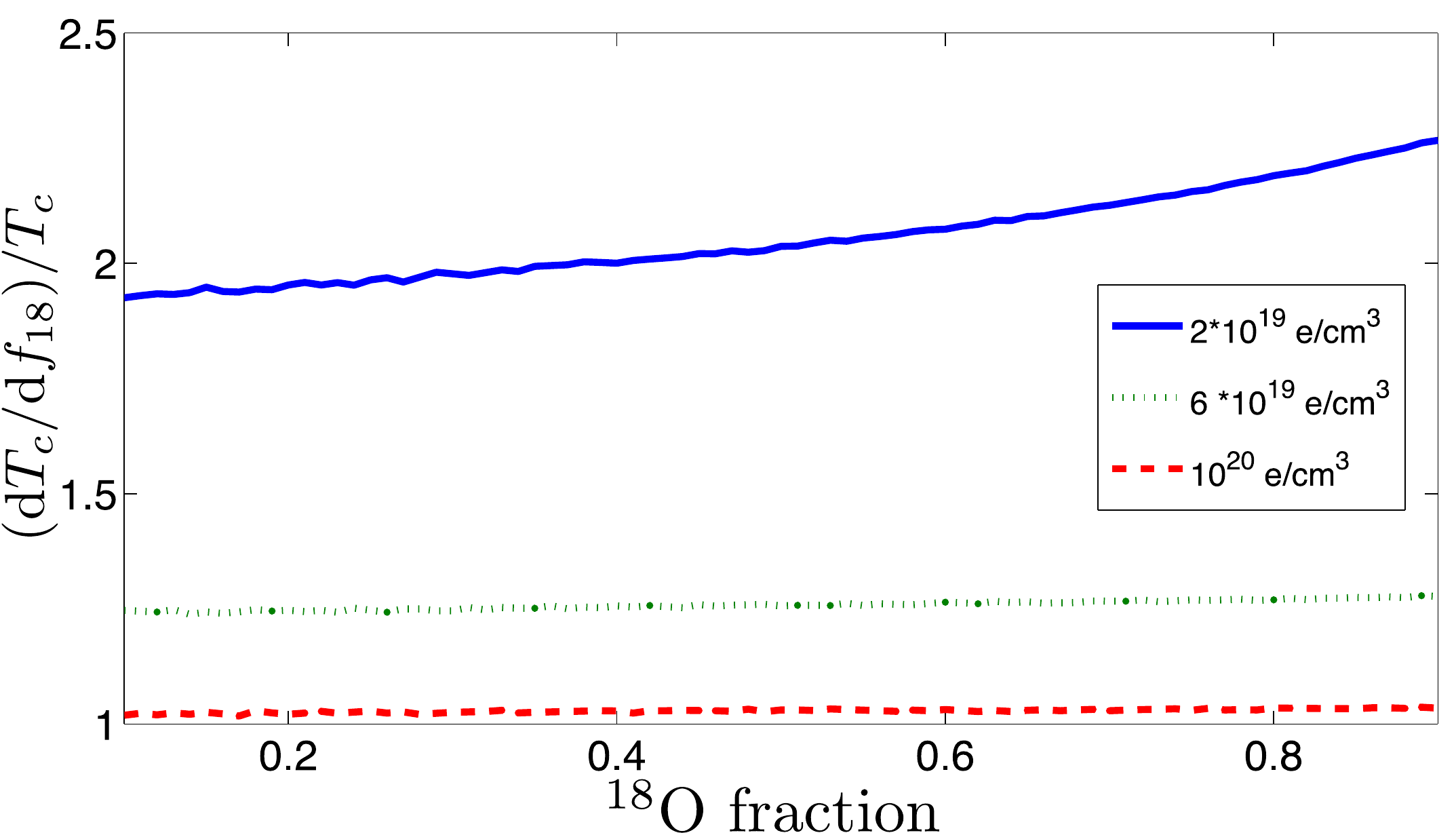}
\caption{The fractional change in the superconducting critical temperature, $\frac1{T_c}\dro{T_c}{f_{18}}$, as a function of $^{18}$O fraction, for several carrier concentrations. The plots are based on the numerical calculations used to produce the superconducting dome in Fig 3 in the main text
}
\label{fig:Dfrac}
\end{figure}

Using Eq.~(7) in the main text, we can calculate the change in $T_c$ as a function of $f_{18}$. We obtain
\begin{align}
\label{eq:1}
\frac1{T_c}\dro{T_c}{f_{18}} &= \frac{\frac A{2\lambda} + \frac{A\alpha^2}{4\lambda^2} \int_{-\pi}^\pi\frac\Gamma{(\Gamma^2 - \Gamma \cos q)^{3/2}} } 
{ \frac{E_F}{\sqrt{T_c} D} \int_{-E_F/T_c}^0 \frac{\tanh(x/2)}{2x \sqrt{x + E_F/T_t}} - \frac1{2\lambda} }
.
\end{align}
A careful study of this expression shows it to be positive for $\Gamma >1$, as both the numerator and denominator are positive. More insight into this function can be gained from its plot, see Fig.~\ref{fig:Dfrac}. Apart from the fact that the change of $T_c$ is positive, we can see that it is larger for lower doping, i.e. when the system is closer to the QCP. This explains the shift of the peak of the dome to lower doping with increasing $^{18}$O fraction. 
For very small doping we can see that $(\dif T_c/\dif f_{18})/T_c$ increases for large $f_{18}$, implying a non-linear dependence on the mass. This suggests that the system is very close to the QCP, in contrast to the case of moderate  and large  doping, where the change is roughly constant for the available mass variation.

\end{widetext}


\begin{thebibliography}{23}%
\makeatletter
\providecommand \@ifxundefined [1]{%
 \@ifx{#1\undefined}
}%
\providecommand \@ifnum [1]{%
 \ifnum #1\expandafter \@firstoftwo
 \else \expandafter \@secondoftwo
 \fi
}%
\providecommand \@ifx [1]{%
 \ifx #1\expandafter \@firstoftwo
 \else \expandafter \@secondoftwo
 \fi
}%
\providecommand \natexlab [1]{#1}%
\providecommand \enquote  [1]{``#1''}%
\providecommand \bibnamefont  [1]{#1}%
\providecommand \bibfnamefont [1]{#1}%
\providecommand \citenamefont [1]{#1}%
\providecommand \href@noop [0]{\@secondoftwo}%
\providecommand \href [0]{\begingroup \@sanitize@url \@href}%
\providecommand \@href[1]{\@@startlink{#1}\@@href}%
\providecommand \@@href[1]{\endgroup#1\@@endlink}%
\providecommand \@sanitize@url [0]{\catcode `\\12\catcode `\$12\catcode
  `\&12\catcode `\#12\catcode `\^12\catcode `\_12\catcode `\%12\relax}%
\providecommand \@@startlink[1]{}%
\providecommand \@@endlink[0]{}%
\providecommand \url  [0]{\begingroup\@sanitize@url \@url }%
\providecommand \@url [1]{\endgroup\@href {#1}{\urlprefix }}%
\providecommand \urlprefix  [0]{URL }%
\providecommand \Eprint [0]{\href }%
\providecommand \doibase [0]{http://dx.doi.org/}%
\providecommand \selectlanguage [0]{\@gobble}%
\providecommand \bibinfo  [0]{\@secondoftwo}%
\providecommand \bibfield  [0]{\@secondoftwo}%
\providecommand \translation [1]{[#1]}%
\providecommand \BibitemOpen [0]{}%
\providecommand \bibitemStop [0]{}%
\providecommand \bibitemNoStop [0]{.\EOS\space}%
\providecommand \EOS [0]{\spacefactor3000\relax}%
\providecommand \BibitemShut  [1]{\csname bibitem#1\endcsname}%
\let\auto@bib@innerbib\@empty
\bibitem [{\citenamefont {Koonce}\ \emph {et~al.}(1967)\citenamefont {Koonce},
  \citenamefont {Cohen}, \citenamefont {Schooley}, \citenamefont {Hosler},\
  and\ \citenamefont {Pfeiffer}}]{Koonce1967}%
  \BibitemOpen
  \bibfield  {author} {\bibinfo {author} {\bibfnamefont {C.~S.}\ \bibnamefont
  {Koonce}}, \bibinfo {author} {\bibfnamefont {M.~L.}\ \bibnamefont {Cohen}},
  \bibinfo {author} {\bibfnamefont {J.~F.}\ \bibnamefont {Schooley}}, \bibinfo
  {author} {\bibfnamefont {W.~R.}\ \bibnamefont {Hosler}}, \ and\ \bibinfo
  {author} {\bibfnamefont {E.~R.}\ \bibnamefont {Pfeiffer}},\ }\href {\doibase
  10.1103/PhysRev.163.380} {\bibfield  {journal} {\bibinfo  {journal} {Phys.
  Rev.}\ }\textbf {\bibinfo {volume} {163}},\ \bibinfo {pages} {380} (\bibinfo
  {year} {1967})}\BibitemShut {NoStop}%
\bibitem [{\citenamefont {Binnig}\ \emph {et~al.}(1980)\citenamefont {Binnig},
  \citenamefont {Baratoff}, \citenamefont {Hoenig},\ and\ \citenamefont
  {Bednorz}}]{Binnig1980}%
  \BibitemOpen
  \bibfield  {author} {\bibinfo {author} {\bibfnamefont {G.}~\bibnamefont
  {Binnig}}, \bibinfo {author} {\bibfnamefont {A.}~\bibnamefont {Baratoff}},
  \bibinfo {author} {\bibfnamefont {H.~E.}\ \bibnamefont {Hoenig}}, \ and\
  \bibinfo {author} {\bibfnamefont {J.~G.}\ \bibnamefont {Bednorz}},\ }\href
  {\doibase 10.1103/PhysRevLett.45.1352} {\bibfield  {journal} {\bibinfo
  {journal} {Phys. Rev. Lett.}\ }\textbf {\bibinfo {volume} {45}},\ \bibinfo
  {pages} {1352} (\bibinfo {year} {1980})}\BibitemShut {NoStop}%
\bibitem [{\citenamefont {Lin}\ \emph {et~al.}(2013)\citenamefont {Lin},
  \citenamefont {Zhu}, \citenamefont {Fauqu\'{e}},\ and\ \citenamefont
  {Behnia}}]{Lin2013}%
  \BibitemOpen
  \bibfield  {author} {\bibinfo {author} {\bibfnamefont {X.}~\bibnamefont
  {Lin}}, \bibinfo {author} {\bibfnamefont {Z.}~\bibnamefont {Zhu}}, \bibinfo
  {author} {\bibfnamefont {B.}~\bibnamefont {Fauqu\'{e}}}, \ and\ \bibinfo
  {author} {\bibfnamefont {K.}~\bibnamefont {Behnia}},\ }\href {\doibase
  10.1103/PhysRevX.3.021002} {\bibfield  {journal} {\bibinfo  {journal} {Phys.
  Rev. X}\ }\textbf {\bibinfo {volume} {3}},\ \bibinfo {pages} {021002}
  (\bibinfo {year} {2013})}\BibitemShut {NoStop}%
\bibitem [{\citenamefont {M\"{u}ller}\ and\ \citenamefont
  {Burkard}(1979)}]{Muller1979}%
  \BibitemOpen
  \bibfield  {author} {\bibinfo {author} {\bibfnamefont {K.}~\bibnamefont
  {M\"{u}ller}}\ and\ \bibinfo {author} {\bibfnamefont {H.}~\bibnamefont
  {Burkard}},\ }\href {\doibase 10.1103/PhysRevB.19.3593} {\bibfield  {journal}
  {\bibinfo  {journal} {Phys. Rev. B}\ }\textbf {\bibinfo {volume} {19}},\
  \bibinfo {pages} {3593} (\bibinfo {year} {1979})}\BibitemShut {NoStop}%
\bibitem [{\citenamefont {Suzuki}\ \emph {et~al.}(2013)\citenamefont {Suzuki},
  \citenamefont {Inoue},\ and\ \citenamefont {Chakrabarti}}]{suzuki2013}%
  \BibitemOpen
  \bibfield  {author} {\bibinfo {author} {\bibfnamefont {S.}~\bibnamefont
  {Suzuki}}, \bibinfo {author} {\bibfnamefont {J.-i.}\ \bibnamefont {Inoue}}, \
  and\ \bibinfo {author} {\bibfnamefont {B.~K.}\ \bibnamefont {Chakrabarti}},\
  }\href {\doibase 10.1007/978-3-642-33039-1_1} {\emph {\bibinfo {title}
  {Quantum Ising Phases and Transitions in Transverse Ising Models}}},\
  \bibinfo {series} {Lecture Notes in Physics}, Vol.\ \bibinfo {volume} {862}\
  (\bibinfo  {publisher} {Springer Berlin Heidelberg},\ \bibinfo {year}
  {2013})\ pp.\ \bibinfo {pages} {1--11}\BibitemShut {NoStop}%
\bibitem [{\citenamefont {Bussmann-Holder}\ \emph {et~al.}(1989)\citenamefont
  {Bussmann-Holder}, \citenamefont {Simon},\ and\ \citenamefont
  {B{\"u}ttner}}]{BussmannHolder/Simon/Buttner:1989}%
  \BibitemOpen
  \bibfield  {author} {\bibinfo {author} {\bibfnamefont {A.}~\bibnamefont
  {Bussmann-Holder}}, \bibinfo {author} {\bibfnamefont {A.}~\bibnamefont
  {Simon}}, \ and\ \bibinfo {author} {\bibfnamefont {H.}~\bibnamefont
  {B{\"u}ttner}},\ }\href@noop {} {\bibfield  {journal} {\bibinfo  {journal}
  {Phys. Rev. B}\ }\textbf {\bibinfo {volume} {39}},\ \bibinfo {pages} {207}
  (\bibinfo {year} {1989})}\BibitemShut {NoStop}%
\bibitem [{\citenamefont {de~Lima}\ \emph {et~al.}(2015)\citenamefont
  {de~Lima}, \citenamefont {da~Luz}, \citenamefont {Oliveira}, \citenamefont
  {Alves}, \citenamefont {dos Santos}, \citenamefont {Jomard}, \citenamefont
  {Sidis}, \citenamefont {Bourges}, \citenamefont {Harms}, \citenamefont
  {Grams}, \citenamefont {Hemberger}, \citenamefont {Lin}, \citenamefont
  {Fauqu{\'e}},\ and\ \citenamefont {Behnia}}]{deLima_et_al:2015}%
  \BibitemOpen
  \bibfield  {author} {\bibinfo {author} {\bibfnamefont {B.~S.}\ \bibnamefont
  {de~Lima}}, \bibinfo {author} {\bibfnamefont {M.~S.}\ \bibnamefont {da~Luz}},
  \bibinfo {author} {\bibfnamefont {F.~S.}\ \bibnamefont {Oliveira}}, \bibinfo
  {author} {\bibfnamefont {L.~M.~S.}\ \bibnamefont {Alves}}, \bibinfo {author}
  {\bibfnamefont {C.~A.~M.}\ \bibnamefont {dos Santos}}, \bibinfo {author}
  {\bibfnamefont {F.}~\bibnamefont {Jomard}}, \bibinfo {author} {\bibfnamefont
  {Y.}~\bibnamefont {Sidis}}, \bibinfo {author} {\bibfnamefont
  {P.}~\bibnamefont {Bourges}}, \bibinfo {author} {\bibfnamefont
  {S.}~\bibnamefont {Harms}}, \bibinfo {author} {\bibfnamefont {C.~P.}\
  \bibnamefont {Grams}}, \bibinfo {author} {\bibfnamefont {J.}~\bibnamefont
  {Hemberger}}, \bibinfo {author} {\bibfnamefont {X.}~\bibnamefont {Lin}},
  \bibinfo {author} {\bibfnamefont {B.}~\bibnamefont {Fauqu{\'e}}}, \ and\
  \bibinfo {author} {\bibfnamefont {K.}~\bibnamefont {Behnia}},\ }\href@noop {}
  {\bibfield  {journal} {\bibinfo  {journal} {Phys. Rev. B}\ }\textbf {\bibinfo
  {volume} {91}},\ \bibinfo {pages} {045108} (\bibinfo {year}
  {2015})}\BibitemShut {NoStop}%
\bibitem [{\citenamefont {Appel}(1969)}]{Appel1969}%
  \BibitemOpen
  \bibfield  {author} {\bibinfo {author} {\bibfnamefont {J.}~\bibnamefont
  {Appel}},\ }\href {\doibase 10.1103/PhysRev.180.508} {\bibfield  {journal}
  {\bibinfo  {journal} {Phys. Rev.}\ }\textbf {\bibinfo {volume} {180}},\
  \bibinfo {pages} {508} (\bibinfo {year} {1969})}\BibitemShut {NoStop}%
\bibitem [{\citenamefont {Rowley}\ \emph {et~al.}(2014)\citenamefont {Rowley},
  \citenamefont {Spalek}, \citenamefont {Smith}, \citenamefont {Dean},
  \citenamefont {Itoh}, \citenamefont {Scott}, \citenamefont {Lonzarich},\ and\
  \citenamefont {Saxena}}]{Rowley2014}%
  \BibitemOpen
  \bibfield  {author} {\bibinfo {author} {\bibfnamefont {S.~E.}\ \bibnamefont
  {Rowley}}, \bibinfo {author} {\bibfnamefont {L.~J.}\ \bibnamefont {Spalek}},
  \bibinfo {author} {\bibfnamefont {R.~P.}\ \bibnamefont {Smith}}, \bibinfo
  {author} {\bibfnamefont {M.~P.~M.}\ \bibnamefont {Dean}}, \bibinfo {author}
  {\bibfnamefont {M.}~\bibnamefont {Itoh}}, \bibinfo {author} {\bibfnamefont
  {J.~F.}\ \bibnamefont {Scott}}, \bibinfo {author} {\bibfnamefont {G.~G.}\
  \bibnamefont {Lonzarich}}, \ and\ \bibinfo {author} {\bibfnamefont {S.~S.}\
  \bibnamefont {Saxena}},\ }\href {\doibase 10.1038/nphys2924} {\bibfield
  {journal} {\bibinfo  {journal} {Nat. Phys.}\ }\textbf {\bibinfo {volume}
  {10}},\ \bibinfo {pages} {10} (\bibinfo {year} {2014})}\BibitemShut {NoStop}%
\bibitem [{\citenamefont {Gegenwart}\ \emph {et~al.}(2008)\citenamefont
  {Gegenwart}, \citenamefont {Si},\ and\ \citenamefont
  {Steglich}}]{Gegenwart2008}%
  \BibitemOpen
  \bibfield  {author} {\bibinfo {author} {\bibfnamefont {P.}~\bibnamefont
  {Gegenwart}}, \bibinfo {author} {\bibfnamefont {Q.}~\bibnamefont {Si}}, \
  and\ \bibinfo {author} {\bibfnamefont {F.}~\bibnamefont {Steglich}},\ }\href
  {\doibase 10.1038/nphys892} {\bibfield  {journal} {\bibinfo  {journal} {Nat.
  Phys.}\ }\textbf {\bibinfo {volume} {4}},\ \bibinfo {pages} {186} (\bibinfo
  {year} {2008})}\BibitemShut {NoStop}%
\bibitem [{\citenamefont {Sachdev}(2000)}]{Sachdev2000}%
  \BibitemOpen
  \bibfield  {author} {\bibinfo {author} {\bibfnamefont {S.}~\bibnamefont
  {Sachdev}},\ }\href {\doibase 10.1126/science.288.5465.475} {\bibfield
  {journal} {\bibinfo  {journal} {Science (80-. ).}\ }\textbf {\bibinfo
  {volume} {288}},\ \bibinfo {pages} {475} (\bibinfo {year}
  {2000})}\BibitemShut {NoStop}%
\bibitem [{\citenamefont {Sebastian}\ \emph {et~al.}(2010)\citenamefont
  {Sebastian}, \citenamefont {Harrison}, \citenamefont {Altarawneh},
  \citenamefont {Mielke}, \citenamefont {Liang}, \citenamefont {Bonn},
  \citenamefont {Hardy},\ and\ \citenamefont {Lonzarich}}]{Sebastian2010}%
  \BibitemOpen
  \bibfield  {author} {\bibinfo {author} {\bibfnamefont {S.~E.}\ \bibnamefont
  {Sebastian}}, \bibinfo {author} {\bibfnamefont {N.}~\bibnamefont {Harrison}},
  \bibinfo {author} {\bibfnamefont {M.~M.}\ \bibnamefont {Altarawneh}},
  \bibinfo {author} {\bibfnamefont {C.~H.}\ \bibnamefont {Mielke}}, \bibinfo
  {author} {\bibfnamefont {R.}~\bibnamefont {Liang}}, \bibinfo {author}
  {\bibfnamefont {D.~A.}\ \bibnamefont {Bonn}}, \bibinfo {author}
  {\bibfnamefont {W.~N.}\ \bibnamefont {Hardy}}, \ and\ \bibinfo {author}
  {\bibfnamefont {G.~G.}\ \bibnamefont {Lonzarich}},\ }\href {\doibase
  10.1073/pnas.0913711107} {\bibfield  {journal} {\bibinfo  {journal} {Proc.
  Natl. Acad. Sci. U. S. A.}\ }\textbf {\bibinfo {volume} {107}},\ \bibinfo
  {pages} {6175} (\bibinfo {year} {2010})}\BibitemShut {NoStop}%
\bibitem [{\citenamefont {Monthoux}\ \emph {et~al.}(1991)\citenamefont
  {Monthoux}, \citenamefont {Balatsky},\ and\ \citenamefont
  {Pines}}]{Monthoux1991}%
  \BibitemOpen
  \bibfield  {author} {\bibinfo {author} {\bibfnamefont {P.}~\bibnamefont
  {Monthoux}}, \bibinfo {author} {\bibfnamefont {A.~V.}\ \bibnamefont
  {Balatsky}}, \ and\ \bibinfo {author} {\bibfnamefont {D.}~\bibnamefont
  {Pines}},\ }\href {\doibase 10.1103/PhysRevLett.67.3448} {\bibfield
  {journal} {\bibinfo  {journal} {Phys. Rev. Lett.}\ }\textbf {\bibinfo
  {volume} {67}},\ \bibinfo {pages} {3448} (\bibinfo {year}
  {1991})}\BibitemShut {NoStop}%
\bibitem [{\citenamefont {Scalapino}(2012)}]{ScalapinoRMP}%
  \BibitemOpen
  \bibfield  {author} {\bibinfo {author} {\bibfnamefont {D.~J.}\ \bibnamefont
  {Scalapino}},\ }\href {\doibase 10.1103/RevModPhys.84.1383} {\bibfield
  {journal} {\bibinfo  {journal} {Rev. Mod. Phys.}\ }\textbf {\bibinfo {volume}
  {84}},\ \bibinfo {pages} {1383} (\bibinfo {year} {2012})}\BibitemShut
  {NoStop}%
\bibitem [{\citenamefont {Roussev}\ and\ \citenamefont
  {Millis}(2001)}]{Roussev2001}%
  \BibitemOpen
  \bibfield  {author} {\bibinfo {author} {\bibfnamefont {R.}~\bibnamefont
  {Roussev}}\ and\ \bibinfo {author} {\bibfnamefont {A.~J.}\ \bibnamefont
  {Millis}},\ }\href {\doibase 10.1103/PhysRevB.63.140504} {\bibfield
  {journal} {\bibinfo  {journal} {Phys. Rev. B}\ }\textbf {\bibinfo {volume}
  {63}},\ \bibinfo {pages} {140504} (\bibinfo {year} {2001})}\BibitemShut
  {NoStop}%
\bibitem [{\citenamefont {Lin}\ \emph {et~al.}(2015)\citenamefont {Lin},
  \citenamefont {Rischau}, \citenamefont {van~der Beek}, \citenamefont
  {Fauque},\ and\ \citenamefont {Behnia}}]{Lin2015}%
  \BibitemOpen
  \bibfield  {author} {\bibinfo {author} {\bibfnamefont {X.}~\bibnamefont
  {Lin}}, \bibinfo {author} {\bibfnamefont {C.~W.}\ \bibnamefont {Rischau}},
  \bibinfo {author} {\bibfnamefont {C.~J.}\ \bibnamefont {van~der Beek}},
  \bibinfo {author} {\bibfnamefont {B.}~\bibnamefont {Fauque}}, \ and\ \bibinfo
  {author} {\bibfnamefont {K.}~\bibnamefont {Behnia}},\ }\href
  {http://arxiv.org/abs/1507.01867} {\ ,\ \bibinfo {pages} {5} (\bibinfo {year}
  {2015})},\ \Eprint {http://arxiv.org/abs/1507.01867} {arXiv:1507.01867}
  \BibitemShut {NoStop}%
\bibitem [{\citenamefont {Aschauer}\ and\ \citenamefont
  {Spaldin}(2014)}]{Aschauer2014}%
  \BibitemOpen
  \bibfield  {author} {\bibinfo {author} {\bibfnamefont {U.}~\bibnamefont
  {Aschauer}}\ and\ \bibinfo {author} {\bibfnamefont {N.~A.}\ \bibnamefont
  {Spaldin}},\ }\href {\doibase 10.1088/0953-8984/26/12/122203} {\bibfield
  {journal} {\bibinfo  {journal} {J. Phys. Condens. Matter}\ }\textbf {\bibinfo
  {volume} {26}},\ \bibinfo {pages} {122203} (\bibinfo {year}
  {2014})}\BibitemShut {NoStop}%
\bibitem [{\citenamefont {Sachdev}(2011)}]{Sachdev2011}%
  \BibitemOpen
  \bibfield  {author} {\bibinfo {author} {\bibfnamefont {S.}~\bibnamefont
  {Sachdev}},\ }\href@noop {} {\emph {\bibinfo {title} {{Quantum Phase
  Transitions}}}},\ \bibinfo {edition} {2nd}\ ed.\ (\bibinfo  {publisher}
  {Cambridge University Press},\ \bibinfo {year} {2011})\BibitemShut {NoStop}%
\bibitem [{\citenamefont {Wang}\ and\ \citenamefont {Itoh}(2001)}]{Wang2001}%
  \BibitemOpen
  \bibfield  {author} {\bibinfo {author} {\bibfnamefont {R.}~\bibnamefont
  {Wang}}\ and\ \bibinfo {author} {\bibfnamefont {M.}~\bibnamefont {Itoh}},\
  }\href {\doibase 10.1103/PhysRevB.64.174104} {\bibfield  {journal} {\bibinfo
  {journal} {Phys. Rev. B}\ }\textbf {\bibinfo {volume} {64}},\ \bibinfo
  {pages} {174104} (\bibinfo {year} {2001})}\BibitemShut {NoStop}%
\bibitem [{\citenamefont {McMillan}(1968)}]{McMillan1968}%
  \BibitemOpen
  \bibfield  {author} {\bibinfo {author} {\bibfnamefont {W.}~\bibnamefont
  {McMillan}},\ }\href {\doibase 10.1103/PhysRev.167.331} {\bibfield  {journal}
  {\bibinfo  {journal} {Phys. Rev.}\ }\textbf {\bibinfo {volume} {167}},\
  \bibinfo {pages} {331} (\bibinfo {year} {1968})}\BibitemShut {NoStop}%
\bibitem [{\citenamefont {Yamada}\ \emph {et~al.}(2004)\citenamefont {Yamada},
  \citenamefont {Todoroki},\ and\ \citenamefont {Miyashita}}]{Yamada2004}%
  \BibitemOpen
  \bibfield  {author} {\bibinfo {author} {\bibfnamefont {Y.}~\bibnamefont
  {Yamada}}, \bibinfo {author} {\bibfnamefont {N.}~\bibnamefont {Todoroki}}, \
  and\ \bibinfo {author} {\bibfnamefont {S.}~\bibnamefont {Miyashita}},\ }\href
  {\doibase 10.1103/PhysRevB.69.024103} {\bibfield  {journal} {\bibinfo
  {journal} {Phys. Rev. B}\ }\textbf {\bibinfo {volume} {69}},\ \bibinfo
  {pages} {024103} (\bibinfo {year} {2004})}\BibitemShut {NoStop}%
\bibitem [{\citenamefont {Abrikosov}\ \emph {et~al.}(1965)\citenamefont
  {Abrikosov}, \citenamefont {Gorkov},\ and\ \citenamefont
  {Dzyaloshinski}}]{AGD-QFT_methods_1965}%
  \BibitemOpen
  \bibfield  {author} {\bibinfo {author} {\bibfnamefont {A.~A.}\ \bibnamefont
  {Abrikosov}}, \bibinfo {author} {\bibfnamefont {L.~P.}\ \bibnamefont
  {Gorkov}}, \ and\ \bibinfo {author} {\bibfnamefont {I.~E.}\ \bibnamefont
  {Dzyaloshinski}},\ }\href@noop {} {\emph {\bibinfo {title} {Quantum Field
  theoretical Methods in Statistical Physics}}}\ (\bibinfo  {publisher}
  {Pergamon Press},\ \bibinfo {year} {1965})\BibitemShut {NoStop}%
\bibitem [{\citenamefont {Maxwell}(1950)}]{isotope}%
  \BibitemOpen
  \bibfield  {author} {\bibinfo {author} {\bibfnamefont {E.}~\bibnamefont
  {Maxwell}},\ }\href {\doibase 10.1103/PhysRev.78.477} {\bibfield  {journal}
  {\bibinfo  {journal} {Phys. Rev.}\ }\textbf {\bibinfo {volume} {78}},\
  \bibinfo {pages} {477} (\bibinfo {year} {1950})}\BibitemShut {NoStop}%
\end{thebibliography}

\begin{thebibliography}{9}%
\makeatletter
\providecommand \@ifxundefined [1]{%
 \@ifx{#1\undefined}
}%
\providecommand \@ifnum [1]{%
 \ifnum #1\expandafter \@firstoftwo
 \else \expandafter \@secondoftwo
 \fi
}%
\providecommand \@ifx [1]{%
 \ifx #1\expandafter \@firstoftwo
 \else \expandafter \@secondoftwo
 \fi
}%
\providecommand \natexlab [1]{#1}%
\providecommand \enquote  [1]{``#1''}%
\providecommand \bibnamefont  [1]{#1}%
\providecommand \bibfnamefont [1]{#1}%
\providecommand \citenamefont [1]{#1}%
\providecommand \href@noop [0]{\@secondoftwo}%
\providecommand \href [0]{\begingroup \@sanitize@url \@href}%
\providecommand \@href[1]{\@@startlink{#1}\@@href}%
\providecommand \@@href[1]{\endgroup#1\@@endlink}%
\providecommand \@sanitize@url [0]{\catcode `\\12\catcode `\$12\catcode
  `\&12\catcode `\#12\catcode `\^12\catcode `\_12\catcode `\%12\relax}%
\providecommand \@@startlink[1]{}%
\providecommand \@@endlink[0]{}%
\providecommand \url  [0]{\begingroup\@sanitize@url \@url }%
\providecommand \@url [1]{\endgroup\@href {#1}{\urlprefix }}%
\providecommand \urlprefix  [0]{URL }%
\providecommand \Eprint [0]{\href }%
\providecommand \doibase [0]{http://dx.doi.org/}%
\providecommand \selectlanguage [0]{\@gobble}%
\providecommand \bibinfo  [0]{\@secondoftwo}%
\providecommand \bibfield  [0]{\@secondoftwo}%
\providecommand \translation [1]{[#1]}%
\providecommand \BibitemOpen [0]{}%
\providecommand \bibitemStop [0]{}%
\providecommand \bibitemNoStop [0]{.\EOS\space}%
\providecommand \EOS [0]{\spacefactor3000\relax}%
\providecommand \BibitemShut  [1]{\csname bibitem#1\endcsname}%
\let\auto@bib@innerbib\@empty
\bibitem [{\citenamefont {Togo}\ \emph {et~al.}(2008)\citenamefont {Togo},
  \citenamefont {Oba},\ and\ \citenamefont {Tanaka}}]{Togo:2008jt}%
  \BibitemOpen
  \bibfield  {author} {\bibinfo {author} {\bibfnamefont {A.}~\bibnamefont
  {Togo}}, \bibinfo {author} {\bibfnamefont {F.}~\bibnamefont {Oba}}, \ and\
  \bibinfo {author} {\bibfnamefont {I.}~\bibnamefont {Tanaka}},\ }\href@noop {}
  {\bibfield  {journal} {\bibinfo  {journal} {Phys. Rev. B}\ }\textbf {\bibinfo
  {volume} {78}},\ \bibinfo {pages} {134106} (\bibinfo {year}
  {2008})}\BibitemShut {NoStop}%
\bibitem [{\citenamefont {Perdew}\ \emph {et~al.}(2008)\citenamefont {Perdew},
  \citenamefont {Ruzsinszky}, \citenamefont {Csonka}, \citenamefont {Vydrov},
  \citenamefont {Scuseria}, \citenamefont {Constantin}, \citenamefont {Zhou},\
  and\ \citenamefont {Burke}}]{Perdew:2008fa}%
  \BibitemOpen
  \bibfield  {author} {\bibinfo {author} {\bibfnamefont {J.~P.}\ \bibnamefont
  {Perdew}}, \bibinfo {author} {\bibfnamefont {A.}~\bibnamefont {Ruzsinszky}},
  \bibinfo {author} {\bibfnamefont {G.}~\bibnamefont {Csonka}}, \bibinfo
  {author} {\bibfnamefont {O.}~\bibnamefont {Vydrov}}, \bibinfo {author}
  {\bibfnamefont {G.}~\bibnamefont {Scuseria}}, \bibinfo {author}
  {\bibfnamefont {L.}~\bibnamefont {Constantin}}, \bibinfo {author}
  {\bibfnamefont {X.}~\bibnamefont {Zhou}}, \ and\ \bibinfo {author}
  {\bibfnamefont {K.}~\bibnamefont {Burke}},\ }\href@noop {} {\bibfield
  {journal} {\bibinfo  {journal} {Phys. Rev. Lett.}\ }\textbf {\bibinfo
  {volume} {100}},\ \bibinfo {pages} {136406} (\bibinfo {year}
  {2008})}\BibitemShut {NoStop}%
\bibitem [{\citenamefont {Kresse}\ and\ \citenamefont
  {Hafner}(1993)}]{Kresse:1993ty}%
  \BibitemOpen
  \bibfield  {author} {\bibinfo {author} {\bibfnamefont {G.}~\bibnamefont
  {Kresse}}\ and\ \bibinfo {author} {\bibfnamefont {J.}~\bibnamefont
  {Hafner}},\ }\href@noop {} {\bibfield  {journal} {\bibinfo  {journal} {Phys.
  Rev. B}\ }\textbf {\bibinfo {volume} {47}},\ \bibinfo {pages} {558} (\bibinfo
  {year} {1993})}\BibitemShut {NoStop}%
\bibitem [{\citenamefont {Kresse}\ and\ \citenamefont
  {Hafner}(1994)}]{Kresse:1994us}%
  \BibitemOpen
  \bibfield  {author} {\bibinfo {author} {\bibfnamefont {G.}~\bibnamefont
  {Kresse}}\ and\ \bibinfo {author} {\bibfnamefont {J.}~\bibnamefont
  {Hafner}},\ }\href@noop {} {\bibfield  {journal} {\bibinfo  {journal} {Phys.
  Rev. B}\ }\textbf {\bibinfo {volume} {49}},\ \bibinfo {pages} {14251}
  (\bibinfo {year} {1994})}\BibitemShut {NoStop}%
\bibitem [{\citenamefont {Kresse}\ and\ \citenamefont
  {Furthmuller}(1996{\natexlab{a}})}]{Kresse:1996vk}%
  \BibitemOpen
  \bibfield  {author} {\bibinfo {author} {\bibfnamefont {G.}~\bibnamefont
  {Kresse}}\ and\ \bibinfo {author} {\bibfnamefont {J.}~\bibnamefont
  {Furthmuller}},\ }\href@noop {} {\bibfield  {journal} {\bibinfo  {journal}
  {Comp. Mater. Sci.}\ }\textbf {\bibinfo {volume} {6}},\ \bibinfo {pages} {15}
  (\bibinfo {year} {1996}{\natexlab{a}})}\BibitemShut {NoStop}%
\bibitem [{\citenamefont {Kresse}\ and\ \citenamefont
  {Furthmuller}(1996{\natexlab{b}})}]{Kresse:1996vf}%
  \BibitemOpen
  \bibfield  {author} {\bibinfo {author} {\bibfnamefont {G.}~\bibnamefont
  {Kresse}}\ and\ \bibinfo {author} {\bibfnamefont {J.}~\bibnamefont
  {Furthmuller}},\ }\href@noop {} {\bibfield  {journal} {\bibinfo  {journal}
  {Phys. Rev. B}\ }\textbf {\bibinfo {volume} {54}},\ \bibinfo {pages} {11169}
  (\bibinfo {year} {1996}{\natexlab{b}})}\BibitemShut {NoStop}%
\bibitem [{\citenamefont {Bl{\"o}chl}(1994)}]{Blochl:1994uk}%
  \BibitemOpen
  \bibfield  {author} {\bibinfo {author} {\bibfnamefont {P.~E.}\ \bibnamefont
  {Bl{\"o}chl}},\ }\href@noop {} {\bibfield  {journal} {\bibinfo  {journal}
  {Phys. Rev. B}\ }\textbf {\bibinfo {volume} {50}},\ \bibinfo {pages} {17953}
  (\bibinfo {year} {1994})}\BibitemShut {NoStop}%
\bibitem [{\citenamefont {Kresse}\ and\ \citenamefont
  {Joubert}(1999)}]{Kresse:1999wc}%
  \BibitemOpen
  \bibfield  {author} {\bibinfo {author} {\bibfnamefont {G.}~\bibnamefont
  {Kresse}}\ and\ \bibinfo {author} {\bibfnamefont {D.}~\bibnamefont
  {Joubert}},\ }\href@noop {} {\bibfield  {journal} {\bibinfo  {journal} {Phys.
  Rev. B}\ }\textbf {\bibinfo {volume} {59}},\ \bibinfo {pages} {1758}
  (\bibinfo {year} {1999})}\BibitemShut {NoStop}%
\bibitem [{\citenamefont {Monkhorst}\ and\ \citenamefont
  {Pack}(1976)}]{Monkhorst:1976ta}%
  \BibitemOpen
  \bibfield  {author} {\bibinfo {author} {\bibfnamefont {H.~J.}\ \bibnamefont
  {Monkhorst}}\ and\ \bibinfo {author} {\bibfnamefont {J.~D.}\ \bibnamefont
  {Pack}},\ }\href@noop {} {\bibfield  {journal} {\bibinfo  {journal} {Phys.
  Rev. B}\ }\textbf {\bibinfo {volume} {13}},\ \bibinfo {pages} {5188}
  (\bibinfo {year} {1976})}\BibitemShut {NoStop}%
\end{thebibliography}
\end{document}